\documentclass[usenatbib,fleqn,useAMS]{mnras}

\usepackage{graphicx}
\usepackage{amsmath}
\usepackage{txfonts}
\usepackage{braket}
\usepackage{cancel}
\usepackage{xcolor}

\renewcommand\le\oldleq
\renewcommand\ge\oldgeq
\renewcommand\pi\upi
\newcommand\pdm\upartial
\newcommand\dm{\mathrm d}

\title[The Potential from the Collisionless Boltzmann Equation]{Charting Galactic Accelerations:
When and How to Extract a Unique Potential from the Distribution Function}

\author[An et al.]
{J.~An$^1$, A.~P.~Naik$^2$, N.~W.~Evans$^3$ and C.~Burrage$^2$
\\$^1$ Center for Theoretical Astronomy, Korea Astronomy \& Space Science Institute, 776 Daedeok-daero, Yuseong-gu, Daejeon 34055, Korea (South);
\\$^2$ School of Physics \& Astronomy, University of Nottingham, University Park, Nottingham NG7~2RD, UK;
\\$^3$ Institute of Astronomy, University of Cambridge, Madingley Rd., Cambridge CB3~0HA, UK.}

\pagerange{\pageref{firstpage}--\pageref{lastpage}}\volume{000}\pubyear{2021}
\date{Accepted 2021 July 9. Received 2021 July 5; in original form 2021 May 29}

\begin{document}
\label{firstpage}
\maketitle

\begin{abstract}
The advent of datasets of stars in the Milky Way with six-dimensional phase-space information makes it possible to construct empirically the distribution function (DF).
Here, we show that the accelerations can be uniquely determined from the DF using the collisionless Boltzmann equation, providing the Hessian determinant of the DF with respect to the velocities is non-vanishing. We illustrate this procedure and requirement with some analytic examples. Methods to extract the potential from datasets of discrete positions and velocities of stars are then discussed. Following Green \& Ting, we advocate the use of normalizing flows on a sample of observed phase-space positions to obtain a differentiable approximation of the DF. To then derive gravitational accelerations, we outline a semi-analytic method involving direct solutions of the over-constrained linear equations provided by the collisionless Boltzmann equation. Testing our algorithm on mock datasets derived from isotropic and anisotropic Hernquist models, we obtain excellent accuracies even with added noise. Our method represents a new, flexible and robust means of extracting the underlying gravitational accelerations from snapshots of six-dimensional stellar kinematics of an equilibrium system.
\end{abstract}

\section{Introduction}

In galactic astronomy, a fundamental problem is to extract the underlying gravitational potential from the kinematics of a tracer population. If stars are moving on circular orbits in a spherical potential, then the matching of the centrifugal force to the gravitational one gives the rotation curve, and by extension the potential. Elaborations of this basic idea to stellar streams have proved to be one of the most powerful methods available to us today \cite[e.g.,][]{Ly82,Jo99,BBE15,Er19,MI19}.

If the stellar population is not kinematically cold, the traditional way in which the problem is tackled is via the Jeans equations \citep[][chap.~4]{BT}. Given measurements of the second velocity moments and the density of the tracer population, the Jeans equations can be solved to yield the potential. There are numerous applications of this method both to the Milky Way \citep[e.g.,][]{Ki15,BEW16,NCN20} and external galaxies \citep[e.g.,][]{Ca08,Wa09}.
Some studies have instead worked directly with the distribution function, fitting some assumed parametric form to the observed stellar data \citep[e.g.,][]{BP15,WE15,PH19}.
More rarely, the distribution function is constructed directly from the data, as in \citet{KG89}'s numerical Abel inversion of the vertical tracer density. This though relies on the assumption that the vertical and in-plane dynamics are decoupled, and so is not of general applicability.

However, the advent of the \textsl{Gaia} satellite \citep{Ga16} has made possible the empirical construction of the full phase-space distribution function for stellar populations in the Milky Way, and perhaps even for some of its satellite galaxies. The data now comprise the full positions and velocities of many millions of stars. The process of averaging to obtain the second velocity moments does not do justice to the richness of the data. \citet{GT20} recently raised the possibility of direct determination of the gravitational potential from the distribution function using the collisionless Boltzmann equation itself. This is the continuity equation satisfied by the distribution function in the six-dimensional phase space of positions and velocities.

At every location in physical space, the collisionless Boltzmann equation provides a \emph{single} constraint on the \emph{three} unknown components of the gravitational force. Thus, it is unclear if the identification of a stationary distribution function is sufficient to specify uniquely the gravitational potential (modulo an additive constant). So, the first aim of our paper is to establish the conditions under which the potential can be uniquely recovered, given the distribution function. The second aim of our paper is to provide a working algorithm to extract the potential. Whereas \citet{GT20} proposed a neural network, we instead utilize an efficient and accurate semi-analytic method, based on a direct solution of the collisionless Boltzmann equation. We demonstrate the efficacy of our method on mock datasets sampled from isotropic and anisotropic distribution functions of galaxy models, including the effects of errors.

\section{The Collisionless Boltzmann Equation and The Potential}
\label{sec:cbe}

Here, we address the theoretical question that underlies all this work: namely, when is the potential uniquely specified by the distribution function? We prove a uniqueness theorem in Section~\ref{sec:uniqueness} subject to certain conditions, and investigate the instances when the conditions are violated in Section~\ref{sec:uniq}. 

\subsection{Uniqueness theorem}
\label{sec:uniqueness}

If $F(\bmath p;\bmath x)$ is a phase-space distribution function (DF) in equilibrium in the static potential $\Phi(\bmath x)$, then it is an integral of motion of the Hamiltonian $H=\frac12\sum_{j,k=1}^3g^{jk}p_jp_k+\Phi(\bmath x)$. Here $\bmath p=(p_1,p_2,p_3)$ is the momentum component conjugate to a coordinate set $\bmath x=(x^1,x^2,x^3)$ with the metric coefficients $g_{ij}$ and its inverse $g^{ij}$. A mathematical representation of $F$ being an integral of motion is given by the vanishing Poisson bracket of the integral $F$ with the Hamiltonian $H$ \citep{AE16}: namely,
\begin{multline}\label{eq:cbe0}
\left\lbrace F,H\right\rbrace
=\sum_{i=1}^3\left(\frac{\pdm F}{\pdm x^i}\frac{\pdm H}{\pdm p_i}
-\frac{\pdm F}{\pdm p_i}\frac{\pdm H}{\pdm x^i}\right)
\\=\sum_{i=1}^3\left[\sum_{j=1}^3g^{ij}p_j\frac{\pdm F}{\pdm x^i}
-\left(\frac12\sum_{j,k=1}^3\frac{\pdm g^{jk}}{\pdm x^i}p_jp_k
+\frac{\pdm\Phi}{\pdm x^i}\right)
\frac{\pdm F}{\pdm p_i}\right]=0.
\end{multline}
Considered as a partial differential equation for $F$,
this is equivalent to the (time-independent) collisionless Boltzmann equation (CBE).\footnote{The CBE is actually derived from arguments based on number (or probability) conservation \citep[see][sect.~4.1]{BT}. In fact, $F$ being an integral of motion may be interpreted as a consequence of the CBE.}
Since equation~(\ref{eq:cbe0}) is a linear homogeneous equation for $F$, any function $F=f(J)$ is its solution if $J=J(\bmath p;\bmath x)$ is also a solution. That is, the CBE only describes a (necessary) condition for the DF to be stationary and cannot uniquely determine the DF for any given potential. In fact, physical considerations make it obvious that many different DFs can indeed be in equilibrium with the given potential.

On the other hand, if a stationary DF is known, the CBE may also be interpreted as a partial differential equation for the potential. Here the question is whether the given DF (or more generally an integral of motion) can determine a unique potential through the CBE. The CBE is linear in $\Phi$ (albeit non-homogeneous) and so there exists a gauge freedom such that, if $\Phi_0$ is a particular solution, the function $\Phi_0+G(I)$, where $G(I)$ is an arbitrary function of a particular solution $I$ to the homogeneous counterpart, also satisfies the same inhomogeneous differential equation. 
However, the potential $\Phi=\Phi(\bmath x)$ is a function of only the configuration-space coordinates, whereas the CBE is a partial differential equation in phase space. In other words, we must only consider the solutions that are also constant along any direction in momentum space; that is, the solution must also be subject to the constraints that $\pdm\Phi/\pdm p_1=\pdm\Phi/\pdm p_2=\pdm\Phi/\pdm p_3=0$. Are these then sufficient to uniquely determine the potential $\Phi$ for the given DF?

Let us suppose that a DF $F(\bmath p;\bmath x)$ is known to be stationary in the potential $\Phi_0(\bmath x)$. Then it follows that $\{F,H_0\}=0$ where $H_0=\sum_{i,j=1}^3\frac12g^{ij}p_ip_j+\Phi_0$ or
\begin{equation}\label{eq:cbe2}
\sum_{i=1}^3\left(\frac{\pdm F}{\pdm x^i}\sum_{j=1}^3 g^{ij}p_j
-\frac12\frac{\pdm F}{\pdm p_i}\sum_{j,k=1}^3\frac{\pdm g^{jk}}{\pdm x^i}p_jp_k
\right)=\sum_{i=1}^3\frac{\pdm\Phi_0}{\pdm x^i}
\frac{\pdm F}{\pdm p_i}.
\end{equation}
If there exists another potential $\Phi$ which the same $F$ is also a stationary DF in, the potential $\Phi$ satisfies the CBE with $F$ in equation~(\ref{eq:cbe0}) or equivalently equation~(\ref{eq:cbe2}) but with $\Phi_0\to\Phi$. Eliminating the common terms between two CBEs, we can construct a homogeneous linear partial differential equation for the difference $\Phi-\Phi_0$:
\begin{equation}\label{eq:lind}
\frac{\pdm(\Phi-\Phi_0)}{\pdm x^1}\frac{\pdm F}{\pdm p_1}
+\frac{\pdm(\Phi-\Phi_0)}{\pdm x^2}\frac{\pdm F}{\pdm p_2}
+\frac{\pdm(\Phi-\Phi_0)}{\pdm x^3}\frac{\pdm F}{\pdm p_3}=0.
\end{equation}
Here $\Phi-\Phi_0$ is a function of only the real-space component $(x^1,x^2,x^3)$, whereas $F$ is a function of phase space in general. Thus taking the partial derivative with respect to one of the momentum components results in the set of three differential equations:
\begin{subequations}
\begin{equation}\label{eq:de}
\sum_{i=1}^3\frac{\pdm(\Phi-\Phi_0)}{\pdm x^i}\frac{\pdm^2F}{\pdm p_j\pdm p_i}=0
\qquad\text{(where $j=1,2,3$)}.
\end{equation}
Since the Hessian matrix $[\pdm_{p_i}\pdm_{p_j}F]$ (where $\pdm_{p_i}=\pdm/\pdm p_i$ and so on) is real symmetric, it is diagonalizable at least locally by a point-wise orthogonal transformation. In the local coordinate diagonalizing the Hessian (in which $\pdm_{\tilde p_i}\pdm_{\tilde p_j}F=0$ for $i\ne j$), equations~(\ref{eq:de}) reduce to
\begin{equation}
\lambda_1\frac{\pdm(\Phi-\Phi_0)}{\pdm q^1}
=\lambda_2\frac{\pdm(\Phi-\Phi_0)}{\pdm q^2}
=\lambda_3\frac{\pdm(\Phi-\Phi_0)}{\pdm q^3}=0
\end{equation}\end{subequations}
Therefore, if $\lambda_i=\pdm_{\tilde p_i}^2F\ne0$ for a direction in the transformed coordinate, then $\pdm(\Phi-\Phi_0)/\pdm q^i=0$ along the conjugate coordinate direction associated with the non-zero eigenvalue $\lambda_i$. If $m$ is the rank (i.e.\ the number of non-zero eigenvalues) of the Hessian, the difference $\Phi-\Phi_0$ is consequently an arbitrary function of $3-m$ functionally-independent functions $q^j=q^j(x^1,x^2,x^3)$, which are the coordinate functions corresponding to the eigenvectors associated with the null eigenvalues.

In particular, if the Hessian determinant
\begin{equation}\label{eq:nvh}
\mbox{det}\left[\frac{\pdm^2F}{\pdm p_i\pdm p_j}\right]
=\mbox{det}\left[\frac{\pdm^2F}{\pdm\tilde p_i\pdm\tilde p_j}\right]
=\lambda_1\lambda_2\lambda_3\ne0
\end{equation}
is non-vanishing, then $\lambda_i\ne0$ for all $i$ and $m=3$. Solving equations~(\ref{eq:de}) as a series of linear equations for $\pdm(\Phi-\Phi_0)/\pdm x^i$ then results in
\begin{equation}
\frac{\pdm(\Phi-\Phi_0)}{\pdm x^1}
=\frac{\pdm(\Phi-\Phi_0)}{\pdm x^2}
=\frac{\pdm(\Phi-\Phi_0)}{\pdm x^3}=0\
\Rightarrow\
\Phi=\Phi_0+C,
\end{equation}
where $C$ is an arbitrary constant; that is, the potential $\Phi(\bmath x)$ satisfying the CBE for a given DF, if it exists, is essentially unique up to an additive constant (resulting in the identical gravitational acceleration field). In other words, the non-vanishing Hessian of equation~(\ref{eq:nvh}) is a sufficient condition for the uniqueness of the potential for a given stationary DF.

\subsection{Are there physical DFs that do not specify a unique potential?}
\label{sec:uniq}

If the Hessian $[\pdm_{p_i}\pdm_{p_j}F]$ is singular, there exists a local momentum-space coordinate system $(\tilde p_1,\tilde p_2,\tilde p_3)$ such that the directional derivative of $F$ in a fixed coordinate direction must be constant in momentum space. That is to say, the singularity condition indicates that at least one eigenvalue, which is the second-order partial derivative in the corresponding coordinate direction, must be zero (i.e.\ $\lambda_j=\pdm_{\tilde p_j}^2F=0$ for $\exists j$). Since the coordinate can be chosen to be orthogonal so that all the second-order cross partial derivatives vanish ($\pdm_{\tilde p_i}\pdm_{\tilde p_j}F=0$ if $i\ne j$), there then exists a coordinate system in which all the second derivatives involving one particular coordinate should be zero (i.e.\ $\pdm_{\tilde p_i}\pdm_{\tilde p_j}F=0$ for $\forall i$ and $\exists j$). Therefore the directional derivative of $F$ in the same coordinate direction must be constant in momentum space; that is, $\pdm_{\tilde p_j}F=k_0(\bmath x)$ for $\exists j$. In the original coordinates, this implies that $\sum_{i=1}^3k_i\pdm_{p_i}F=k_0$ where $k_i$'s are the constants in momentum space (but they are functions of the real-space positions) and at least one of $\set{k_1(\bmath x),k_2(\bmath x),k_3(\bmath x)}$ is nonzero. In fact, if there are two or more distinct potentials satisfying the CBE with the given DF, equation~(\ref{eq:lind}) further indicates that there exists $\set{k_1,k_2,k_3}$ such that $k_1^2+k_2^2+k_3^2\ne0$ and $(k_1\pdm_{p_1}+k_2\pdm_{p_2}+k_3\pdm_{p_3})F=0$.

In other words, if the function $F$ is an integral of motion in two (or more) distinct -- as in generating different gravitational accelerations -- potentials, then there exists a fixed direction $(k_1,k_2,k_3)$ in momentum space that is tangent to the level surfaces of the DF everywhere in momentum space. However, the integral curve of a constant vector is a straight line and momentum space is topologically equivalent to $\mathbb R^3$. Consequently, all the level surfaces of $F$ have infinite extent and the inverse image of any real interval under $F^{-1}$ in momentum space cannot have a compact support (unless empty). That is to say, such a function $F$ is not integrable and cannot be a physical DF.

In light of this, we argue that the unique determination of the potential is a property related to the global behaviour in momentum space. That is to say, the CBE only describes the balance amongst the gradients of the DF and the external acceleration field in the local neighborhood of a fixed phase-space location, whilst the external gravitational acceleration is shared in the whole momentum space at a fixed real-space position. By joining all the constraints on the acceleration fields coming from the CBE in different momentum-space locations (but at a fixed real-space position), we can narrow down to the unique acceleration. This fact is also demonstrated by the examples presented in the following section (Sect.~\ref{sec:ex}) where a unique potential actually follows from insisting that the CBE holds for all values of the momentum components.

\section{Examples}
\label{sec:ex}

To gain insight into the steps needed to extract a unique potential from the CBE, we first look at some analytic examples.

\subsection{Ergodic distributions: a unique potential}

We start by examining the case of an ergodic DF $F=f(E)$ in a fixed potential $\Phi_0(\bmath x)$, where $E=\frac12\varv^2+\Phi_0$ is the specific energy and is known as a function of the phase-space coordinates. Here, no further assumption is made on the self-consistency of the system and so the potential need not be spherically symmetric \citep[cf.][]{AES17}.
In Cartesian coordinates, the CBE is then reducible to the differential equation on the difference between any two possible potentials:
\begin{multline}
\left(\varv_x\frac{\pdm E}{\pdm x}
+\varv_y\frac{\pdm E}{\pdm y}
+\varv_z\frac{\pdm E}{\pdm z}
-\frac{\pdm\Phi}{\pdm x}\frac{\pdm E}{\pdm\varv_x}
-\frac{\pdm\Phi}{\pdm y}\frac{\pdm E}{\pdm\varv_y}
-\frac{\pdm\Phi}{\pdm z}\frac{\pdm E}{\pdm\varv_z}\right)f'(E)
\\=f'(E)\left(\varv_x\frac\pdm{\pdm x}
+\varv_y\frac\pdm{\pdm y}+\varv_z\frac\pdm{\pdm z}\right)(\Phi_0-\Phi)=0.
\end{multline}
Assuming that the DF in itself is not constant, that is, $f'(E)\ne0$, then in order for this to hold everywhere in phase space,
\begin{equation}
\frac{\pdm(\Phi-\Phi_0)}{\pdm x}
=\frac{\pdm(\Phi-\Phi_0)}{\pdm y}
=\frac{\pdm(\Phi-\Phi_0)}{\pdm z}=0.
\end{equation}
Therefore $\Phi=\Phi_0+C$ and the potential is unique (up to an additive constant).

\subsection{Separable potentials with third integrals}

If there exists a DF of the form $F=f(J)$ where $f'(J)\ne0$ and $J$ is a quadratic polynomial of $p_i$'s such that $J=\sum_{i,j=1}^3\frac12K^{ij}(\bmath x)p_ip_j+\sum_{i=1}^3X^i(\bmath x)p_i+\xi(\bmath x)$, then the resulting CBE in equation~(\ref{eq:cbe0}) reduces to a cubic polynomial equation on $p_i$'s. This is of course the old ``ellipsoidal hypothesis'' \citep[see][and references therein]{Ch39,Ca41,EL91}. Assuming that the DF is stationary, the CBE should hold for any $p_i$'s and so the coefficients to all the monomial terms ($p_ip_jp_k$, $p_ip_j$, and $p_i$ etc.) must vanish identically. It is then found that the coefficients to the cubic and quadratic terms respectively only involve the tensor $K^{ij}$ and the vector $X^i$, and the first-order partial differential equations resulting from setting them to be zero restrict the possible forms for $K^{ij}$ and $X^i$ \citep[and references therein]{An13}. However if the DF is already given and known to be stationary, these conditions must hold automatically.

On the other hand, setting the coefficients to the linear terms to be zero results in the set of three differential equations:
\begin{equation}\label{eq:sepp}
\sum_{i=1}^3g^{ij}\frac{\pdm\xi}{\pdm x^i}
=\sum_{i=1}^3K^{ij}\frac{\pdm\Phi}{\pdm x^i}
\qquad\text{(where $j=1,2,3$)}.
\end{equation}
If $\xi(\bmath x)$ is known, these can be considered as the coupled differential equations on the potential $\Phi$. Provided that the matrix $[K^{ij}]$ is invertible (here also note that $K^{ij}=\pdm_{p_i}\pdm_{p_j}J$), equation~(\ref{eq:sepp}) can be uniquely solved for $\pdm\Phi/\pdm x^i$ so that
\begin{equation}
\frac{\pdm\Phi}{\pdm x^i}=\sum_{j=1}^3K^{-1}_{ij}
\left(\sum_{k=1}^3g^{jk}\frac{\pdm\xi}{\pdm x^k}\right)
\qquad\text{(where $i=1,2,3$)},
\end{equation}
where $K^{-1}_{ij}$ is the matrix element of the inverse matrix of $[K^{ij}]$. In other words, if the local DF that is a function of a non-degenerate quadratic form of the canonical momenta is stationary, the gravitational acceleration is uniquely specified in the neighborhood.

As a concrete example, suppose that there exists a stationary DF of the form $F=f(J)$ where
\begin{equation}
J=\frac{\ell^2+a^2\varv_z^2}2+\xi(R,z);\quad
\xi=\frac{ka|z|}{[R^2+(|z|+a)^2]^{1/2}}
\end{equation}
with constants $a$ and $k$, is the third integral of the \citet{Ku56} disc potential in the cylindrical polar coordinate $(R,\phi,z)$ and $\ell=\lVert\bmath\ell\rVert=(\bmath{\ell\cdot\ell})^{1/2}$ is the magnitude of the specific angular momentum.
Here, $\bmath\ell=\bmath x\bmath\times\dot{\bmath x}=(R\bmath{\hat e}_R+z\bmath{\hat e}_z)\bmath\times(\varv_R\bmath{\hat e}_R+\varv_\phi\bmath{\hat e}_\phi+\varv_z\bmath{\hat e}_z)$ and so follows that
$\ell^2=(z\varv_R-R\varv_z)^2+(R^2+z^2)\varv_\phi^2$, whilst $(p_R,p_\phi,p_z)=(\varv_R,R\varv_\phi,\varv_z)$.
%
%
Provided $f'(J)\ne0$, the CBE in the corresponding canonical phase-space coordinate $(p_R,p_\phi,p_z;R,\phi,z)$ then results in (here $r^2=R^2+z^2$)
\begin{multline}
p_R\frac{\pdm\xi}{\pdm R}+p_z\frac{\pdm\xi}{\pdm z}
+z(Rp_z-zp_R)\frac{\pdm\Phi}{\pdm R}
\\\shoveright{
+[Rzp_R-(R^2+a^2)p_z]\frac{\pdm\Phi}{\pdm z}
-\frac{r^2p_\phi}{R^2}\frac{\pdm\Phi}{\pdm\phi}
}\\\shoveleft{
=p_R\left(\frac{\pdm\xi}{\pdm R}-z^2\frac{\pdm\Phi}{\pdm R}
+Rz\frac{\pdm\Phi}{\pdm z}\right)
}\\
+p_z\left[\frac{\pdm\xi}{\pdm z}+Rz\frac{\pdm\Phi}{\pdm R}
-(R^2+a^2)\frac{\pdm\Phi}{\pdm z}\right]
-\frac{r^2p_\phi}{R^2}\frac{\pdm\Phi}{\pdm\phi}=0.
\end{multline}
Since this holds for all $(p_R,p_\phi,p_z)$, we have $\pdm\Phi/\pdm\phi=0$ and
\begin{subequations}\label{eq:xide}\begin{gather}
z^2\frac{\pdm\Phi}{\pdm R}
-Rz\frac{\pdm\Phi}{\pdm z}
=\frac{\pdm\xi}{\pdm R}
=-\frac{kaR|z|}{[R^2+(|z|+a)^2]^{3/2}};
\\
(R^2+a^2)\frac{\pdm\Phi}{\pdm z}
-Rz\frac{\pdm\Phi}{\pdm R}
=\frac{\pdm\xi}{\pdm z}
=\frac z{|z|}\frac{ka[R^2+a(|z|+a)]}{[R^2+(|z|+a)^2]^{3/2}},
\end{gather}\end{subequations}
where we have used $\pdm|z|/\pdm z=z/|z|$ (NB: $\pdm\xi/\pdm z$ at $z=0$ does not exist). If $a\ne0$, we can solve equation~(\ref{eq:xide}) for $\pdm\Phi/\pdm R$ and $\pdm\Phi/\pdm z$:
\begin{subequations}\label{eq:pcde}\begin{gather}
\frac{\pdm\Phi}{\pdm R}
=\frac1{a^2}\left(\frac{R^2+a^2}{z^2}\frac{\pdm\xi}{\pdm R}
+\frac Rz\frac{\pdm\xi}{\pdm z}\right)
=\frac{kR}{[R^2+(|z|+a)^2]^{3/2}};
\\
\frac{\pdm\Phi}{\pdm z}
=\frac1{a^2}\left(\frac Rz\frac{\pdm\xi}{\pdm R}+\frac{\pdm\xi}{\pdm z}\right)
=\frac z{|z|}\frac{k(|z|+a)}{[R^2+(|z|+a)^2]^{3/2}},
\end{gather}\end{subequations}
which satisfies the compatibility condition, $\pdm_z(\pdm_R\Phi)=\pdm_R(\pdm_z\Phi)$. Since $\Phi=\Phi(R,z)$ which follows $\pdm\Phi/\pdm\phi=0$, equation~(\ref{eq:pcde}) can be directly integrated to yield a unique solution:
\begin{equation}
\Phi=-\frac k{[R^2+(|z|+a)^2]^{1/2}}+C,
\end{equation}
which recovers the axisymmetric potential of the Kuzmin disc up to an additive constant $C$.

\subsection{Integrals of motion due to the symmetry of the potential}

Let us consider the DF $F=f(\ell_z)$ where $\ell_z=\bmath{\ell\cdot\hat e_z}$ is the component of the specific angular momentum in a fixed (say, Cartesian $z$) direction. Technically any such function cannot be integrable over the whole phase space and so is unphysical. Nevertheless, the CBE merely requires $F$ to be an integral of motion, and so is still applicable. Since $\ell_z=R^2\dot\phi=R\varv_\phi$, the CBE in phase-space coordinates $(\varv_R,\varv_\phi,\varv_z;R,\phi,z)$ thus simplifies to
\begin{subequations}\begin{multline}
\dot R\frac{\pdm F}{\pdm R}
+\dot\varv_\phi\frac{\pdm F}{\pdm\varv_\phi}
=\left[\varv_R\frac{\pdm \ell_z}{\pdm R}
-\frac1R\left(\varv_R\varv_\phi+\frac{\pdm\Phi}{\pdm\phi}\right)
\frac{\pdm \ell_z}{\pdm\varv_\phi}\right]f'(\ell_z)
\\=\left[\cancel{\varv_R\varv_\phi}-
\left(\cancel{\varv_R\varv_\phi}+\frac{\pdm\Phi}{\pdm\phi}\right)\right]f'(\ell_z)
=-\frac{\pdm\Phi}{\pdm\phi}f'(\ell_z)=0.
\end{multline}
Or given that $\ell_z=p_\phi$, the CBE in the canonical phase-space coordinate $(p_R,p_\phi,p_z;R,\phi,z)$ simply becomes
\begin{equation}
\dot p_\phi\frac{\pdm F}{\pdm p_\phi}
=-\frac{\pdm\Phi}{\pdm\phi}f'(\ell_z)=0.
\end{equation}\end{subequations}
Provided $f'(\ell_z)\ne0$, this implies that $\pdm\Phi/\pdm\phi=0$, the general solution of which is any axisymmetric potential; that is, an arbitrary function $\Phi=\Phi(R,z)$ of two coordinate functions $R$ and $z$. Also note $\pdm F/\pdm p_R=\pdm F/\pdm p_z=0$ indicates that the only non-zero component of the Hessian $[\pdm_{p_i}\pdm_{p_j}F]$ is $\pdm^2F/\pdm p_\phi^2$ and so it follows that the rank of Hessian is 1 as long as $\pdm^2F/\pdm p_\phi^2=f''(\ell_z)\ne0$.

The result is independent of the choice of the coordinate, although the calculation may be more complicated. For example, in Cartesian coordinates, $\ell_z=x\varv_y-y\varv_x$ and so the Hessian becomes
\begin{equation}
\begin{bmatrix}\pdm_{\varv_x}^2F&
\pdm_{\varv_x}\pdm_{\varv_y}F&
\pdm_{\varv_x}\pdm_{\varv_z}F\\
\pdm_{\varv_y}\pdm_{\varv_x}F&
\pdm_{\varv_y}^2F&
\pdm_{\varv_y}\pdm_{\varv_z}F\\
\pdm_{\varv_z}\pdm_{\varv_x}F&
\pdm_{\varv_z}\pdm_{\varv_y}F&
\pdm_{\varv_z}^2F\end{bmatrix}
=f''(\ell_z)\begin{bmatrix}y^2&-xy&0\\-xy&x^2&0\\0&0&0\end{bmatrix}
\end{equation}
whose rank is still 1 if $f''(\ell_z)\ne0$. The CBE on the other hand is
\begin{subequations}\begin{equation}
\left(\cancel{\varv_x\varv_y-\varv_y\varv_x}+\frac{\pdm\Phi}{\pdm x}y
-\frac{\pdm\Phi}{\pdm y}x\right)f'(\ell_z)=0
\end{equation}
and so, unless $f'(\ell_z)=0$, we have a homogeneous first-order linear partial differential equation on $\Phi(x,y,z)$,
\begin{equation}
\frac{\pdm\Phi}{\pdm x}y-\frac{\pdm\Phi}{\pdm y}x=0.
\end{equation}\end{subequations}
Utilizing standard techniques such as the method of characteristics, its general solution is found to be $\Phi=\Phi(x^2+y^2,z)$, which is again an arbitrary axisymmetric function.

Similarly if a stationary DF (or rather an integral of motion) of the form $F=f(\ell^2)$ is available, the CBE in the canonical phase-space coordinate $(p_r,p_\theta,p_\phi;r,\theta,\phi)$ inherited from the spherical polar coordinate $(r,\theta,\phi)$ is reducible to
\begin{subequations}\begin{multline}
\left[\frac{p_\theta}{r^2}\frac{\pdm \ell^2}{\pdm\theta}
+\left(\frac{p_\phi^2\cos\theta}{r^2\sin^3\theta}
-\frac{\pdm\Phi}{\pdm\theta}\right)\frac{\pdm \ell^2}{\pdm p_\theta}
-\frac{\pdm\Phi}{\pdm\phi}\frac{\pdm \ell^2}{\pdm p_\phi}\right]f'(\ell^2)
\\=2\left[-\frac{p_\theta}{r^2}\frac{p_\phi^2\cos\theta}{\sin^3\theta}
+\left(\frac{p_\phi^2\cos\theta}{r^2\sin^3\theta}
-\frac{\pdm\Phi}{\pdm\theta}\right)p_\theta
-\frac{\pdm\Phi}{\pdm\phi}\frac{p_\phi}{\sin^2\theta}\right]f'(\ell^2)
\\=-2\left(p_\theta\frac{\pdm\Phi}{\pdm\theta}
+\frac{p_\phi}{\sin^2\theta}\frac{\pdm\Phi}{\pdm\phi}\right)f'(\ell^2),
\end{multline}
which follows as $\ell^2=r^2(\varv_\theta^2+\varv_\phi^2)=p_\theta^2+p_\phi^2/(\sin^2\theta)$ and $(p_\theta,p_\phi)=(r\varv_\theta,r\varv_\phi\sin\theta)$.
%
%
Assuming $f'(\ell^2)\ne0$, this is equivalent to
\begin{equation}\label{eq:cbess}
p_\theta\frac{\pdm\Phi}{\pdm\theta}
+\frac{p_\phi}{\sin^2\theta}\frac{\pdm\Phi}{\pdm\phi}
=r\left(\varv_\theta\frac{\pdm\Phi}{\pdm\theta}
+\frac{\varv_\phi}{\sin\theta}\frac{\pdm\Phi}{\pdm\phi}\right)=0.
\end{equation}\end{subequations}
If $f(\ell^2)$ is a non-constant integral of motion, equation~(\ref{eq:cbess}) should hold everywhere in phase space (i.e.\ for any $p_\theta$ and $p_\phi$) and so
\begin{equation}
\frac{\pdm\Phi}{\pdm\theta}=\frac{\pdm\Phi}{\pdm\phi}=0\
\Rightarrow\
\Phi=\Phi(r).
\end{equation}
Hence the general solution is any spherically symmetric potential.
As for the rank of the corresponding Hessian, we observe that the rank of the matrix $[\pdm_{p_i}\pdm_{p_j}\ell^2]$ is 2 (independent of the coordinate system) with the radial vector being the eigenvector associated with a null eigenvalue (note $\pdm \ell^2/\pdm p_r=0$ in the spherical polar coordinate). In addition the radial vector is also in the null space of the matrix $[(\pdm_{p_i}\ell^2)(\pdm_{p_j}\ell^2)]$, thanks again to $\pdm \ell^2/\pdm p_r=0$. Hence, for any $f(\ell^2)$, the radial vector is in the null space of the Hessian matrix;
\begin{equation}
\left[\frac{\pdm^2F}{\pdm p_i\pdm p_j}\right]
=f'(\ell^2)\left[\frac{\pdm^2\ell^2}{\pdm p_i\pdm p_j}\right]
+f''(\ell^2)\left[\left(\frac{\pdm \ell^2}{\pdm p_i}\right)
\left(\frac{\pdm \ell^2}{\pdm p_j}\right)\right].
\end{equation}
In other words, the Hessian is singular and its rank is at most 2.

Since any axisymmetric or spherical potential admits the integral of motion $\ell_z$ or $\ell^2$, it is not an unexpected result that $F=f(\ell_z)$ or $f(\ell^2)$ only constrains the associated symmetry of the potential and cannot specify the unique potential. The above examples however demonstrate that such integrals of motion also fail the necessary condition of having a non-singular Hessian in momentum space. Furthermore, we also observe that $f(\ell_z)$ and $f(\ell^2)$ are not actually integrable in momentum space. That is to say, $f(\ell_z)$ is independent of $p_R$ and $p_z$, but both components are unbounded, and so the integral of any non-negative $f(\ell_z)$ over the whole momentum space is infinite (unless it is identically zero). A similar argument can also be made for $f(\ell^2)$ and the component $p_r$. We have argued in Section~\ref{sec:uniq} that this is not an accident, but that there is a logical connection between the singular Hessian and the non-integrability.

\section{Algorithms for extracting the gravitational acceleration}

Suppose that stationary DF $F$ is known. How then can we extract the gravitational accelerations? First consider the CBE in an arbitrary curvilinear orthogonal coordinate (in which the line element is $\dm s^2=h_1^2\dm x_1^2+h_2^2\dm x_2^2+h_3^2\dm x_3^2$) rearranged to be
\begin{subequations}\begin{gather}\label{eq:cbe1}
\frac{\pdm F}{\pdm\varv_1}\frac{\pdm\Phi}{h_1\pdm x_1}
+\frac{\pdm F}{\pdm\varv_2}\frac{\pdm\Phi}{h_2\pdm x_2}
+\frac{\pdm F}{\pdm\varv_3}\frac{\pdm\Phi}{h_3\pdm x_3}=S;
\\\label{eq:sdef}
S=\sum_{i=1}^3\varv_i\frac{\pdm F}{h_i\pdm x_i}
+\sum_{i,j=1}^3\varv_j\left(\varv_j\frac{\pdm\ln|h_j|}{h_i\pdm x_i}
-\varv_i\frac{\pdm\ln|h_i|}{h_j\pdm x_j}\right)
\frac{\pdm F}{\pdm\varv_i},
\end{gather}\end{subequations}
where $\varv_i=h_i\dot x_i$ is the velocity component projected on the orthonormal frame.
Next let us observe that $\bmath\nabla\Phi$ is constant at all the velocity-space points, given a fixed real-space position. Hence the subset of equations~(\ref{eq:cbe1}) sampled over the range of velocity space at a fixed position results in an over-determined (assuming there are more than three sampling points) system of linear equations on $(\pdm\Phi/\pdm x_1,\pdm\Phi/\pdm x_2,\pdm\Phi/\pdm x_3)$. Technically, we only need samples at three different velocity-space points so as to uniquely determine the local gravitational acceleration, provided that the three vectors $\bmath{\nabla_\varv}\,F$ at the three sampled points -- where $\bmath{\nabla_\varv}=(\pdm_{\varv_1},\pdm_{\varv_2},\pdm_{\varv_3})$ is the gradient operator in velocity space -- are mutually linearly independent. In fact, the non-singular Hessian of $F$ as discussed in Section~\ref{sec:cbe} guarantees the existence of such three points in velocity space (and so is a sufficient condition for the unique determination of the potential).

On physical grounds, the over-determined system of equations~(\ref{eq:cbe1}) resulting from more than three velocity-space points at a single spatial location should be consistent and must possess a unique solution. However, due to the uncertainties in the data, the exact solution may not be necessarily found with the actual set of equations in practice. Instead, the problem should be approached by methods such as least-square: that is, minimizing
\begin{equation}\label{eq:norm}
\sum_{\text{sample}}\frac1{\varsigma^2}\left(
\frac{\pdm F}{\pdm\varv_1}\frac{\pdm\Phi}{h_1\pdm x_1}
+\frac{\pdm F}{\pdm\varv_2}\frac{\pdm\Phi}{h_2\pdm x_2}
+\frac{\pdm F}{\pdm\varv_3}\frac{\pdm\Phi}{h_3\pdm x_3}
-S\right)^2
\end{equation}
where $S$ is as defined in equation~(\ref{eq:sdef}), and the summation is over a suitably-chosen sample of velocities with the weights $\varsigma^{-2}$. Finding the extrema with respect to $\bmath\nabla\Phi=(\pdm\Phi/\pdm x_1,\pdm\Phi/\pdm x_2,\pdm\Phi/\pdm x_3)$ is then equivalent to solving the set of linear equations:
\begin{subequations}\label{eq:extle}\begin{gather}
\sum_{i=1}^3A_{ij}\frac{\pdm\Phi}{h_i\pdm x_i}=
\sum_{\text{sample}}\frac S{\varsigma^2}
\frac{\pdm F}{\pdm\varv_j}
\qquad\text{(where $j=1,2,3$)},
\\\label{eq:cmat}\text{where }\
A_{ij}=\sum_{\text{sample}}\frac1{\varsigma^2}
\frac{\pdm F}{\pdm\varv_i}\frac{\pdm F}{\pdm\varv_j},
\end{gather}\end{subequations}
which is basically the set of standard normal equations. Therefore, provided the matrix $[A_{ij}]$ defined as in equation~(\ref{eq:cmat}) is invertible, $\bmath\nabla\Phi$ that minimizes equation~(\ref{eq:norm}) at the same position can be found through a matrix inversion.

Alternatively one may also attempt to minimize equation~(\ref{eq:norm}) summed over data points ranging in a region of space, in order to get the potential as an optimizing functional solution. In principle, this can be done with a suitably-chosen parametric function for the potential or non-parametrically (pixelized or otherwise), which is closer to the implementation proposed by \citet{GT20} to recover the potential. After reconstructing the DF from the discrete dataset via normalizing flows, \citet{GT20} characterized the potential as an optimized feed-forward neural network minimizing the cost function, which is defined similarly to equation~(\ref{eq:norm}) but with the absolute value instead of the square and also includes the penalty for the negative density. This procedure combines the determination of the local accelerations and their integration into the potential as one single optimization problem. Nonetheless, the actual physical constraints due to the CBE are in the form of an algebraic relation on the local acceleration and so the measurements of the accelerations at different spatial locations should in principle be independent (except for possible systematic correlations relating to the determination of the DF).

\section{Effects of disequilibrium}

If the stellar system is not in equilibrium, its DF $F(\bmath p;\bmath x;t)$ by definition, is no longer an integral of motion. Provided that the collisional effects are negligible, the evolution of the DF is still governed by the CBE, but the CBE now must include explicit time dependence; $D_tF=\pdm_tF+\{F,H\}=0$, where $D_tF$ is the (Lagrangian) phase-space convective derivative and $\pdm_tF=\pdm F/\pdm t$ is the (Eulerian) time rate of change of $F$ at a fixed phase-space coordinate, whilst $\{F,H\}$ is the same as equation~(\ref{eq:cbe0}). We observe that the argument in Section~\ref{sec:uniqueness} still holds for the time-dependent CBE as long as $\pdm_tF$ is also a known quantity. In particular, equation~(\ref{eq:cbe1}) maintains the same form but the right-hand side additionally includes the $\pdm_tF$ term ($S\to S+\pdm_tF$), and so the determination of the acceleration is still possible if $\pdm_tF$'s are known throughout phase space. However $\pdm_tF$ is impossible to measure directly within a practical time scale barring few exceptional situations -- by contrast, if $\bmath\nabla\Phi$ is known independently, $\pdm_tF$ may instead be determined using the CBE. If $\pdm_tF$ is considered as unknown, the system of equations (\ref{eq:cbe1}) becomes under-constrained and the problem is technically insoluble without some additional restrictive assumptions on the behaviours of $\bmath\nabla\Phi$ or $\pdm_tF$.

Nevertheless, we may still infer effects due to the system not being in equilibrium. If the time derivatives are neglected when not warranted, that will introduce a systematic bias. Notably, the linear system of equations~(\ref{eq:cbe1}) would then not necessarily be consistent even if all the phase-space derivatives of $F$ are known exactly. Whilst equation~(\ref{eq:extle}) still has a unique solution despite the system of equations~(\ref{eq:cbe1}) being inconsistent, the resulting solution is actually offset by the ``sample average'' of $\pdm_tF$. That is to say, if $\pdm\Phi^\mathrm s/\pdm x_i$ is the solution of inverting equation~(\ref{eq:extle}) with $\pdm_tF=0$ (whereas $\pdm\Phi/\pdm x_i$ is the true gravitational acceleration component), then
\begin{equation}\label{eq:bias}
\frac{\pdm\Phi^\mathrm s}{h_i\pdm x_i}=\frac{\pdm\Phi}{h_i\pdm x_i}-\sum_{j=1}^3A^{-1}_{ij}T_j,\
\text{where }\
T_j=\sum_{\text{sample}}\frac1{\varsigma^2}
\frac{\pdm F}{\pdm\varv_j}\frac{\pdm F}{\pdm t}
\end{equation}
and $A^{-1}_{ij}$ is the matrix element of the inverse matrix of $[A_{ij}]$ in equation~(\ref{eq:cmat}). This follows from the fact that $\pdm\Phi/\pdm x_i$ is actually the solution of equation~(\ref{eq:extle}) with $S\to S+\pdm_tF$. If we insert back the solution (eq.~\ref{eq:bias}) into equation~(\ref{eq:cbe1}) and consider the departure from the equality at each sample point, then (with $B_i=\sum_{j=1}^3A^{-1}_{ij}T_j$)
\begin{equation}
\sum_{i=1}^3\frac{\pdm\Phi^\mathrm s}{h_i\pdm x_i}\frac{\pdm F}{\pdm\varv_i}-S
=\frac{\pdm F}{\pdm t}-\sum_{i=1}^3B_i\frac{\pdm F}{\pdm\varv_i}.
\end{equation}
In other words, the residuals consist of the time derivative $\pdm_tF$ and the projection of the bias (i.e.\ $B_i$) onto $\bmath{\nabla_\varv}F$. We note that $B_i$'s are unknown but fixed constants and so the last term is also considered as $\bmath{\nabla_\varv}F$ projected onto a fixed (albeit unknown) direction, which behaves in a predictable systematic pattern. Consequently it would be a smoking gun for a system in disequilibria if the observed residual on each sample point exhibits a systematic behaviour over velocity space not consistent with a projection of $\bmath{\nabla_\varv}F$ onto a fixed direction.

\begin{figure*}
    \centering
    \includegraphics{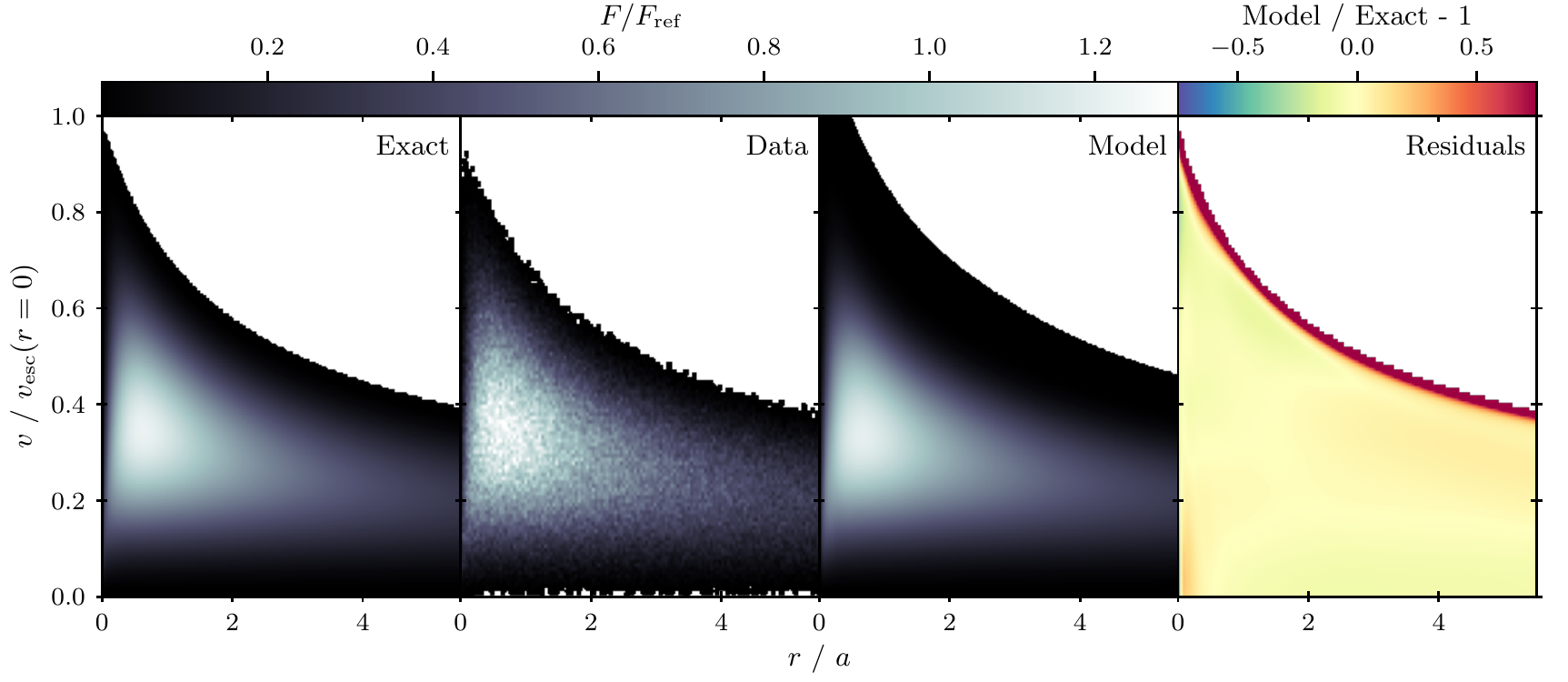}
    \caption{The isotropic Hernquist DF ($M=10^{10}~\mbox{M}_{\sun}, a=5~\mbox{kpc}$), projected into $r$-$\varv$ space. Note that the absolute values of the DF are not of immediate interest, so in each case we have divided by a reference value, given by the \emph{exact} DF evaluated at $r=a$, $\varv = 0.5\varv_\mathrm{esc}(r=a)$. \textit{Left:} the exact DF given by eq.~(\ref{eq:hq_df}). \textit{Second panel:} a histogram of our mock dataset. \textit{Third panel:} the normalizing flow reconstruction of the DF. \textit{Right:} fractional residuals comparing the reconstructed and exact DFs. This figure illustrates that the normalizing flow technique successfully learns the isotropic Hernquist DF.}
    \label{fig:iso}
\end{figure*}

\begin{figure}
    \centering
    \includegraphics{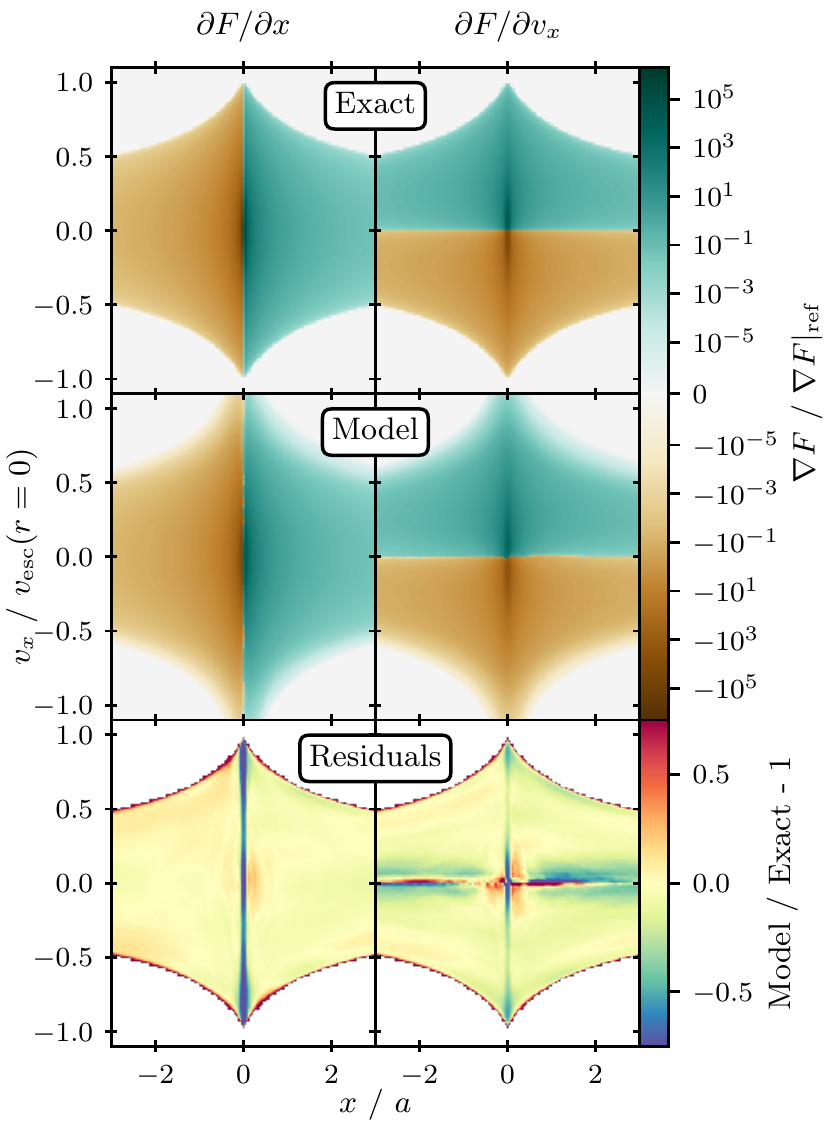}
    \caption{First derivatives of the isotropic Hernquist DF. The left column of three panels shows the spatial derivative $\pdm F / \pdm x$, whilst the right column gives the velocity derivatives $\pdm F / \pdm \varv_x$. The derivatives are everywhere divided by a reference value, evaluated as in Fig.~\ref{fig:iso}. \textit{Top row:} exact derivatives computed by differentiating eq.~(\ref{eq:hq_df}). \textit{Middle row:} derivatives of the flow-reconstructed DF. \textit{Bottom row:} fractional residuals. This figure demonstrates that the normalizing flow technique not only recovers the DF but also its gradients, which are required for measuring accelerations, cf.\ eq.~(\ref{eq:extle}).}
    \label{fig:derivs}
\end{figure}

\begin{figure}
    \centering
    \includegraphics{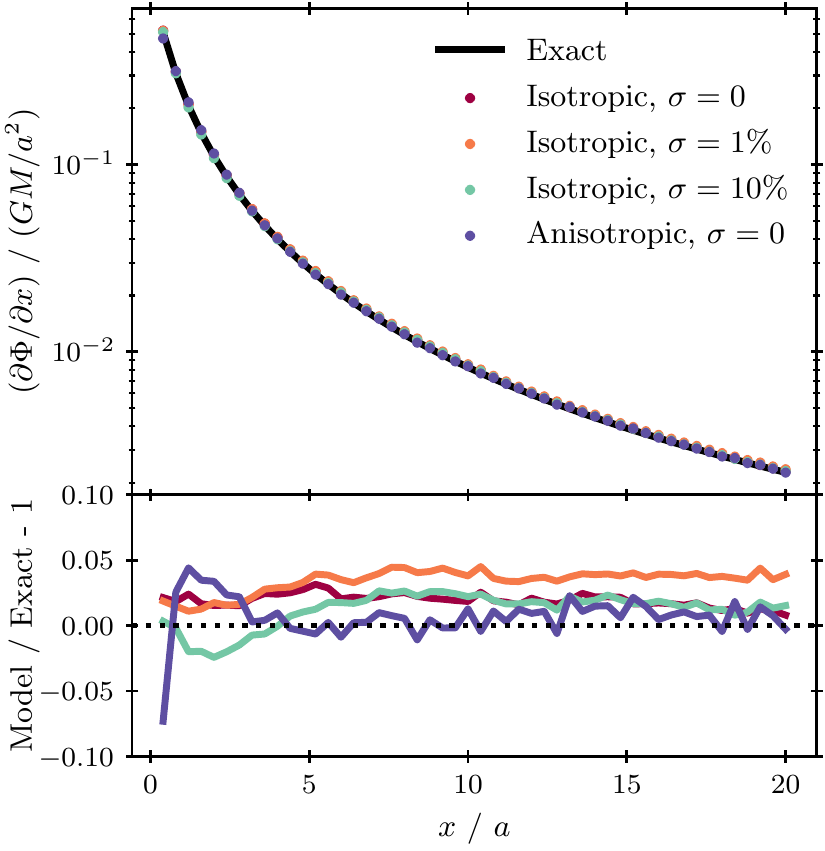}
    \caption{Accelerations in the Hernquist model. The solid black line in the upper panel shows the exact accelerations along the $x$-axis, whilst the points show the accelerations derived by applying eq.~(\ref{eq:extle}) to the non-parametric DF learned by the normalizing flows. The different colours show results for flows trained on different datasets, as labelled in the legend. The lower panel shows fractional residuals. This figure illustrates that our method successfully derives accelerations from a 6D snapshot of kinematic data.}
    \label{fig:accs}
\end{figure}

\begin{figure*}
    \centering
    \includegraphics{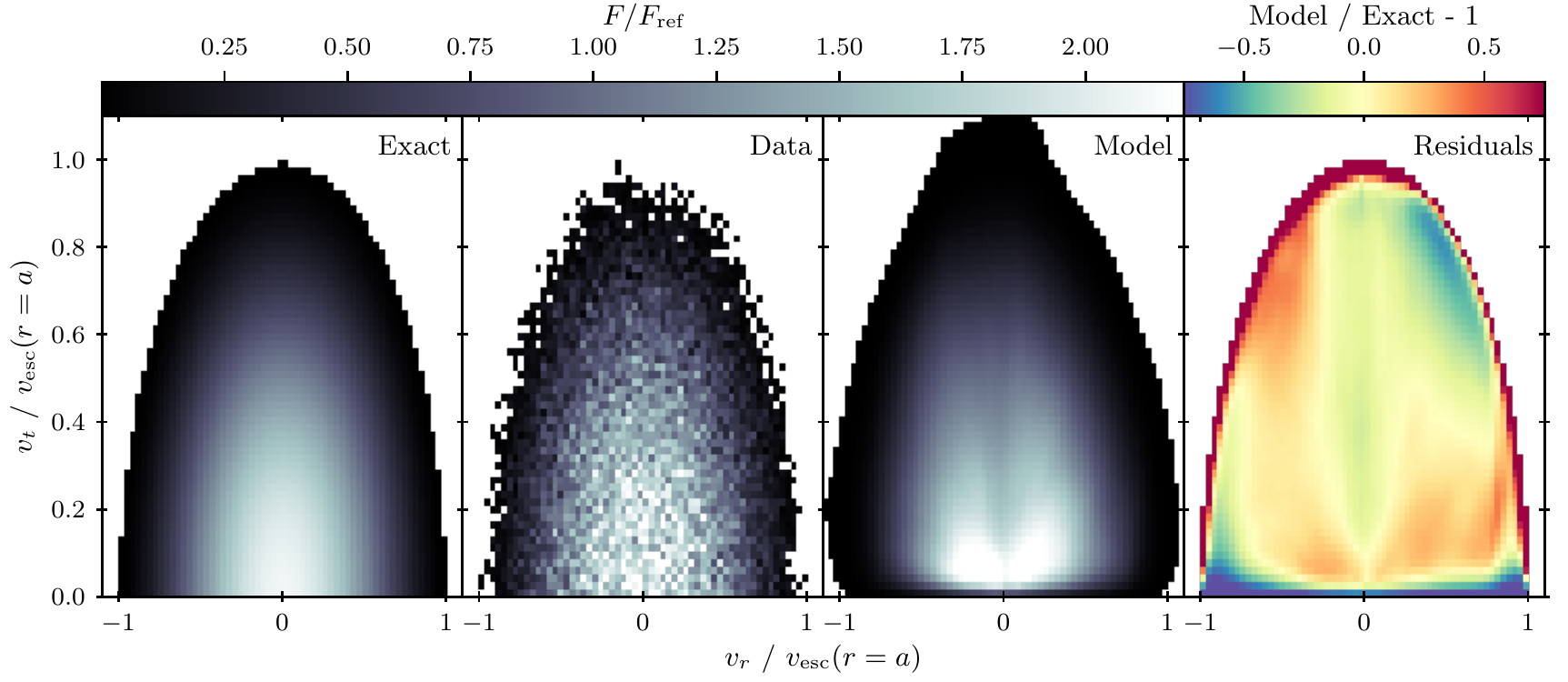}
    \caption{The anisotropic Hernquist DF, projected into $\varv_r$-$\varv_t$ space at fixed position ($r=a$). The four panels carry the same meanings as in their isotropic analogues in Fig.~\ref{fig:iso}, although some differences are discussed in the text. The normalizing flow technique is also successful at recovering the anisotropic Hernquist DF, albeit with larger residuals than in the isotropic case.}
    \label{fig:aniso}
\end{figure*}

\section{Implementation}
\label{sec:implement}

Given a known DF, equation~(\ref{eq:extle}) furnishes us with a way to calculate gravitational accelerations, under the assumption of equilibrium. We now wish to test this technique on a mock dataset.

Here, we demonstrate a complete pipeline from a six-dimensional (6D) stellar kinematics dataset to a map of accelerations. This will necessitate an additional step in the procedure, that is, obtaining the underlying DF of the data. Whereas a conventional approach might have involved fitting the data with a well-motivated analytic DF \citep[e.g.,][]{BP15,WE15,PH19}, we instead follow the philosophy of \citet{GT20} and construct a non-parametric DF directly from the data. 

Our method can thus be summarized:
\begin{enumerate}
    \item Employing a normalizing flow technique, we reconstruct a non-parametric DF from the mock data. 
    \item With this reconstructed DF in hand, we exploit eq.~(\ref{eq:extle}) to calculate accelerations.
\end{enumerate}


This exercise serves mainly as a proof of concept. In a subsequent paper (Naik et al., in prep.), we shall apply the same methodology to local stellar kinematics, with a view towards mapping the acceleration field (and thence the distribution of matter) in the solar neighbourhood.

It is worth noting that an acceleration field calculated with our method is not guaranteed to be physical, in the sense that it might show negative divergences (i.e.\
negative mass densities) or non-zero curls (i.e.\
non-conservative force). We view this feature as an advantage: the existence of such non-Newtonian accelerations can serve as a valuable \textit{post hoc} test of our method. If they are found to be robust, they might hint at disequilibrium features or non-gravitational force (even modified gravity). On the other hand, the requirements for non-negative divergences and vanishing curls can be imposed \textit{a priori} if so desired, by adding penalty terms to the loss function used to train the normalizing flow. These non-Newtonian accelerations are then still possible in principle, but heavily suppressed.

\subsection{Ergodic models}

We consider a simple galaxy halo model in which the DF self-consistently generates both the potential and the density. We generate a mock 6D dataset using this DF, and then attempt to derive the underlying acceleration field from the mock data. For this model, we adopt the spherical \citet{He90} profile, specified by the potential-density pair
\begin{equation}
\label{E:SatPotential}
        \Phi(r) = -\frac{GM}{r+a}; \qquad
        \rho(r) = \frac{M}{2\pi} \frac{a}{r(r+a)^3},
\end{equation}
where $M$ and $a$ are respectively the galaxy mass and scale radius. The isotropic (ergodic) DF for this model is given by\footnote{Here the normalization uses $\int\!\dm^3\!\bmath x\,\dm^3\!\bmath\varv\,F=1$ \citep[cf.][eq.~4.1]{BT}, whereas eq.~(17) of \citet{He90} follows $\int\!\dm^3\!\bmath x\,\dm^3\!\bmath\varv\,F=M$.}
\begin{multline}
\label{eq:hq_df}
    F = f(E) = \frac{1}{8\!\sqrt{2}\pi^3 (GMa)^{3/2}}
    \\ \times \left[ \frac{3\sin^{-1}\!\sqrt{\epsilon}}{(1-\epsilon)^{5/2}} + \frac{\!\sqrt{\epsilon}\,(1-2\epsilon) (8\epsilon^2-8\epsilon-3)}{(1-\epsilon)^2} \right].
\end{multline}
where $\epsilon = -Ea/GM\ge0$ (here, $-E$ is the specific binding energy of a star). In this case, the phase-space gradients of $F$ are determined solely by the gradients of the energy $E$.

A visualization of the isotropic DF, for $M=10^{10}~\mbox{M}_{\sun}$ and $a=5~\mbox{kpc}$, is given in the left-hand panel of Figure~\ref{fig:iso}. There is a clear curve above which the DF is everywhere zero: viz.\ the escape velocity $\varv_\mathrm{esc} = \sqrt{ 2GM / (r+a) }$. With this DF, we employ an MCMC technique to sample a mock 6D dataset with $10^6$ stars. For this, we use the affine-invariant ensemble sampler implemented in the software package \textsc{emcee} \citep{Fo13}. A density plot of this mock dataset is shown in the second panel of Figure~\ref{fig:iso}.

From this mock dataset, we now want to learn the underlying DF by means of a normalizing flow technique \citep{RM15}. Normalizing flows are a relatively new probability density estimation technique, and the basic principle behind them is rather straightforward: a simple base distribution such as a Gaussian is subject to a series (or ``flow'') of complex (but bijective and invertible) transformations into a target distribution. The parameters of these transformations are then optimized so as to give a target distribution that closely resembles the data. More detailed descriptions of the technique are given in the article by \citet{RM15} first describing normalizing flows, and the recent review articles by \citet{KPB20} or \citet{Pa21}.

Despite taking a single Gaussian as the starting point, a flow with sufficiently flexible transformations (and sufficiently many of them) is able to mimic arbitrarily complex, multimodal data distributions. In practice, even rather minimalist flow architectures are capable of achieving great complexity \citep[see e.g.,][for an impressive application of flows in image generation]{Ki18}.

Another class of density estimation technique capable of emulating arbitrarily complex datasets is kernel density estimation. The advantages of flow-based techniques over kernel-based techniques are two-fold. First, flows are less susceptible to over/under-fitting data \citep{BK19}. The second advantage is more context-dependent. Kernel-based techniques typically require no training beyond simply loading the kernels into memory, and perhaps some tuning of the kernel-width parameter. However, given a dataset of size $N$, evaluating the kernel density PDF then essentially requires the computation of $N$ kernel functions, which can be costly as $N$ grows large. Flows do require a training procedure, the cost and duration of which depend on the flow architecture and the size and complexity of the dataset in question. However, given a trained flow, evaluating the PDF is then a mere matter of computing a single Gaussian and a small number of transformations, regardless of $N$. In summary, kernel densities are cheap to train but expensive to evaluate, while flow densities are expensive to train but cheap to evaluate. In our context, we need to train a density estimator only once to learn the DF, but would then like to evaluate it many times, e.g., for the sums in equation~(\ref{eq:extle}). This would therefore suggest flows over kernels.

Another notable aspect of normalizing flows is that the target distribution is guaranteed to be a well-behaved probability distribution, i.e.\ positive everywhere and normalized to unity. The positivity requirement is met straightforwardly by working in log-space, but the normalization requirement is more exacting: it restricts the space of usable transformations to bijective and invertible functions. This space is then restricted further by the desire for computational efficiency.
Different normalizing flow techniques differ primarily in the details of these transformations, as well as the base distributions and flow architectures. 

We differ from \citet{GT20} in that we employ ``masked autoregressive flows'' \citep[MAFs;][]{PPM17}. This choice is motivated by the benchmarking of a number of normalizing flow algorithms. We train an ensemble of 30 MAFs, each with 8 transformations along the flow, each transformation being a neural network with one hidden layer of 64 units. We use the implementation of MAFs in the publicly available software package \textsc{nflows}.\footnote{\textsc{nflows}: normalizing flows in \textsc{PyTorch}, doi:10.5281/zenodo.4296287}

The MAFs are trained on the mock data, and thus learn a non-parametric DF that closely resembles the data. This learned DF is shown in the third panel of Figure~\ref{fig:iso}. It is worth emphasizing that, whilst this plot is in two dimensions, the MAFs are trained using 6D data and learn a 6D DF. The plotted values here are taken from a 2D slice through this 6D DF, with $y=z=\varv_y=\varv_z=0$ (so that $x=r$, $\varv_x=\varv$). The rightmost panel of Figure~\ref{fig:iso} shows fractional residuals, i.e.\ $F_\mathrm{model} / F_\mathrm{exact} - 1$. Encouragingly, the residuals are less than 5~\% throughout most of phase space. In other words, our algorithm is successfully able to reproduce the isotropic Hernquist DF.

One apparent qualification to this success is the region near the $\varv_\mathrm{esc}$-curve, where the DF is consistently overestimated. The $\varv_\mathrm{esc}$-curve represents a hard edge in the Hernquist DF, and even very flexible non-parametric density estimation schemes can struggle to reproduce such a hard edge. However, this need not be a cause for concern, for the following reason: if we progress to step (ii) of our method and attempt to derive acceleration at a given spatial location using this learned DF, the right-hand side of equation~(\ref{eq:extle}) requires us to choose a number of points in velocity space. At this stage, we are free to choose whichever velocities we like, and we can thus choose to steer well clear of this region near $\varv_\mathrm{esc}$, which we term a ``zone of avoidance''. Of course, in real-world applications, one might not know the exact value of $\varv_\mathrm{esc}$, but one can always make an educated guess \citep[e.g.,][]{Wi17,De19}.

Equation~(\ref{eq:extle}) requires the spatial and velocity derivatives of the DF to calculate accelerations. We therefore check if our technique accurately recovers not just the DF, but also its derivatives. Here, a compelling benefit of the normalizing flow technique is that the learned DF is everywhere exactly differentiable, irrespective of the complexity of the flow architecture. Thus, we can efficiently calculate exact derivatives, obviating the need for potentially noisy finite difference schemes.

Figure~\ref{fig:derivs} compares the first derivatives $\pdm F / \pdm x$ and  $\pdm F / \pdm \varv_x$ of the exact and reconstructed DFs, evaluated on a 2D $(x, \varv_x)$ plane in phase space. Inspecting the residuals in the lower panels of Figure~\ref{fig:derivs}, it is apparent that the MAFs are rather successful at accurately recovering the gradients of the DF; the residuals are less than 10~\% throughout most of phase space. 

As seen in Figure~\ref{fig:iso}, there is a problematic region of larger residuals near $\varv_\mathrm{esc}$. In addition to this, two more such regions are apparent. First, the $\pdm F / \pdm \varv_x$ residuals grow rather large in the immediate vicinity of $\varv_x=0$. This is the peak of the 1D $\varv_x$-distribution, and so the nearby gradients are small and susceptible to mis-estimation. Second, the $\pdm F / \pdm x$ residuals show similar issues around $x=0$. The same arguments hold here, perhaps exacerbated by the power-law cusp in the Hernquist model. For calculating accelerations, the first problem can be avoided as in the $\varv_\mathrm{esc}$ case, i.e.\ by sampling velocities that avoid the region around $\varv_x = 0$ (likewise $\varv_y, \varv_z$). However, in the second region around $x=0$, the residuals appear to be consistently large throughout velocity space, suggesting that our calculated accelerations at these small very radii will be biased.

With these points in mind, we now progress to step (ii) of our method, and derive accelerations from our learned DF using equation~(\ref{eq:extle}). Here, we take 50 points along the $x$-axis, and at each of these points we sample $10^3$ velocities for the sums on the right-hand side of equation~(\ref{eq:extle}). We perform this sampling by calculating the escape speed $\varv_\mathrm{esc}$ at each spatial point, then uniformly sampling $10^4$ speeds between 0 and $0.9\,\varv_\mathrm{esc}$. Random directions are then chosen from the unit sphere. Finally, we randomly subsample $10^3$ velocities from this set, avoiding the region around $\varv_i = 0$.

After performing this sampling, we have $10^3$ points in phase space at which we evaluate equation~(\ref{eq:extle}) for each spatial location. The results of this are shown as the ``Isotropic, $\sigma = 0$'' curve in Figure~\ref{fig:accs}. It is clear that the method derives the accelerations in the isotropic Hernquist model very well. The fractional residuals shown in the lower panel indicate an accuracy everywhere at the level of $\la3~\%$.

\subsection{Anisotropic models}

We repeat this exercise, using a simple anisotropic DF for the Hernquist model \citep{BD02,EA05,EA06}
\begin{equation}
\label{E:HernquistDFAniso}
    F = f(E,\ell) = \frac{3}{4 \pi^3 GMa} \frac{\epsilon^2}{\ell}.
\end{equation}
Now, the DF depends on the magnitude of the angular momentum $\ell = r\varv_{\rm t}$ (here $\varv_{\rm t}^2 = \varv_\theta^2 + \varv_\phi^2$) as well as the (dimensionless) binding energy $\epsilon$. As before, we sample one million positions and velocities from this DF, then feed this data to an ensemble of MAFs.

Figure~\ref{fig:aniso} is the anisotropic analogue of Figure~\ref{fig:iso}, and shows the exact DF, a density plot of the mock data, the learned DF and the fractional residuals. As the DF is not isotropic, we do not show the DFs projected into $(r,\varv)$ space, but rather into $(\varv_r,\varv_{\rm t})$ or radial versus tangential velocity space at fixed position ($r=a$). Consequently, the ``Data'' panel does not show the full dataset as in Figure~\ref{fig:iso}, but only the stars within a small radial slice around $r=a$. 

The residuals in the anisotropic DF are generally larger than in the isotropic case, but nonetheless reasonably small, $\sim5$--10~\%. Moreover, there seems to a be an additional zone of avoidance here beyond those already discussed in the isotropic case, around $\varv_{\rm t}=0$. The source of the large errors here can be seen directly from the form of the DF (\ref{E:HernquistDFAniso}): $\varv_{\rm t}=0$ means $\ell=0$, so the DF diverges. The probability distribution remains well-behaved, but the MAFs nonetheless struggle to reproduce the sharp rise in probability density at small $\varv_{\rm t}$.

Despite these foibles, the accelerations are still well recovered in the anisotropic case. These are shown as the points labelled ``Anisotropic, $\sigma=0$'' in Figure~\ref{fig:accs}. Indeed, the residuals here are comparable to the isotropic case.

One aspect of our procedure worth emphasizing is that successful calculation of accelerations relies on the judicious choice of velocity samples, steering clear of the ``zones of avoidance'' in which the DF and its gradients are poorly estimated. We have seen above that the existence and locations of zones can vary from context to context, and so it might be difficult to know \emph{a priori} where they are for any given real stellar population. This is a potential drawback to our method, but it can be readily circumvented by performing tests on mock datasets.

\subsection{Effect of errors}

As a final test, we assess the potential impact of observational errors by adding Gaussian noise to the isotropic dataset, at the 1~\% and 10~\% level. The results of this trial are also shown in Figure~\ref{fig:accs}, alongside the original results for the noiseless dataset. Based on this test, it appears that random errors of this magnitude have no appreciable adverse impact on the calculation of accelerations, with residuals still at the percent level. The application of our method to real data is therefore unlikely to be limited by statistical error.

Going beyond our simple test, there is a natural way to propagate observational errors in our method: when training an ensemble of MAFs on the data, each MAF could be provided with a slightly different dataset from which to learn, generated from a different realisation of the error distribution. Each member of the ensemble will then have a different learned opinion about the acceleration at a given spatial location, and the spread of these values will incorporate observational errors.

\section{Conclusions}

The phase-space distribution (DF) for the stars in the Milky Way is an obvious way to organize the new datasets comprising of nearby stars in the full six-dimensional phase-space coordinates. One question that follows is what information the DF actually contains about the overall properties of the Galaxy. We have proved that, if the stationary DF of a population is known locally in the neighborhood of a fixed real-space position, then the gravitational acceleration at that location can be uniquely determined from the phase-space gradients of the DF, using the collisionless Boltzmann equation (CBE) under the assumption of dynamical equilibrium.
A sufficient condition for this to be true is that the Hessian of the DF with respect to the momenta does not vanish (see eq.~\ref{eq:nvh}).

In practice, once the CBEs are set up locally at more than three independent phase-space points sharing the real-space coordinates, we have an over-determined system of linear equations on the potential gradients, which can be solved via techniques, such as the least square and normal equations. A practical prescription of how to do this is provided in equation~(\ref{eq:extle}).

In light of this finding, we address the question as to how to empirically reconstruct the DF suitable for the local measurements of the gravitational acceleration. Recent developments in machine learning techniques offer great promise in this regard. In particular, \citet{GT20} proposed that the DF of stars can be reconstructed from samples of discrete positions and velocities via the method of normalizing flows and the underlying potential can be recovered from this empirical DF. We examine this suggestion by devising tests derived from isotropic and anisotropic Hernquist models using masked autoregressive flows to build the DF. Once built, direct solution of the over-constrained linear equations for the accelerations (eq.~\ref{eq:extle}) is highly efficient, and preferable to use of a neural network \citep[cf.][]{GT20}. The accelerations are everywhere well reproduced with samplings of $\sim 1000$ velocities at any given position. One caveat here is the existence of regions of velocity space in which the DF is poorly estimated, which need to be avoided in the sampling. Tests with the addition of Gaussian noise at the 1~\% or 10~\% level suggest that the method is stable against errors of this magnitude. 

There are a number of evident applications of this method, some of which we are actively pursuing. For example, if we reconstruct the velocity distributions of a homogeneous (in equilibrium) stellar population in the solar neighbourhood from the sample of the nearby stars \citep[e.g.,][]{Ga20}, it is possible to measure the local gravity at the sun's position due to the Galactic potential (Naik et al., in prep.). This has implications both for the measurement of the local dark matter density and for tests of alternative theories of gravity. Equally, the method is potentially applicable to the datasets of Milky Way halo stars to measure the mass of the Milky Way and its escape speed.

One assumption underlying the implementation of our method is that of dynamical equilibrium. Incorrectly assuming $\pdm F/\pdm t = 0$ leads to an additive bias in the derived accelerations that is linear in $\pdm F/\pdm t$. In addition, disequilibrium can manifest itself through the system of equations~(\ref{eq:cbe1}) sampled at many different velocity space positions being inconsistent with a single value of $\pdm\Phi/\pdm x_i$ (after accounting for observational uncertainties), or equation~(\ref{eq:extle}) resulting in different values of the acceleration for distinct choices of samples. 

There is now a significant body of evidence suggesting the existence of disequilibria in the Milky Way disc \citep[e.g.,][]{Ant18,SD18,Sa20}, which will need to be carefully considered in future applications of our technique to local stellar kinematics. \citet{BWD17} find the bias in inferred accelerations to be at the 10~\% level if such systematic perturbations are ignored. So, it is interesting to explore whether the pattern of residuals at a sampling point has a systematic behaviour over velocity space that may be a tell-tale signature of departures from equilibrium \citep[cf.][for a somewhat similar idea]{LW21}.

It is also worth remarking that the first step of our outlined procedure, i.e.\ learning the DF with normalizing flows, is entirely assumption-free. Given this learned DF, one could then study the non-equilibrium structures themselves. These non-equilbrium structures imprinted in the stellar kinematics are much more than merely sources of systematic error: perturbations to a system can reveal insights about the system itself. For example, \citet{Wi21} has shown that the shape of the Gaia phase spiral can be used to constrain the local gravitational potential.

To summarize, our method bypasses many of the assumptions that have been traditionally adopted in studies of galactic dynamics, and represents an efficient, flexible, and data-driven means of extracting underlying gravitational accelerations from snapshots of stellar kinematics.

\section*{acknowledgement}
We thank Gregory Green and Yuan-Sen Ting for useful discussions, as well as the anonymous referee for a very useful report.
APN and CB are supported by a Research Leadership Award from the Leverhulme Trust. CB is also supported by a Royal Society University Research Fellowship. 

\section*{Data Availability}
The code, mock datasets, and models used in this paper have all been made publicly available at \url{https://github.com/aneeshnaik/HernquistFlows}.

\label{lastpage}
\end{document}